\documentclass[twocolumn,aps,showpacs,preprintnumbers]{revtex4}
\usepackage{amssymb,graphicx}

\begin{document}
\title{Compatibility between shape equation and boundary conditions of lipid membranes with free edges}

\author{Z. C. Tu}\email[E-mail: ]{tuzc@bnu.edu.cn}
\affiliation{Department of Physics, Beijing Normal University,
Beijing 100875, China} \affiliation{Kavli Institute for Theoretical
Physics China, CAS, Beijing 100190, China}

\begin{abstract}
Only some special open surfaces satisfying the shape equation of
lipid membranes can be compatible with the boundary conditions. As a
result of this compatibility, the first integral of the shape
equation should vanish for axisymmetric lipid membranes, from which
two theorems of non-existence are verified: (i) There is no
axisymmetric open membrane being a part of torus satisfying the
shape equation; (ii) There is no axisymmetric open membrane being a
part of a biconcave discodal surface satisfying the shape equation.
Additionally, the shape equation is reduced to a second-order
differential equation while the boundary conditions are reduced to
two equations due to this compatibility. Numerical solutions to the
reduced shape equation and boundary conditions agree well with the
experimental data [A. Saitoh \emph{et al.}, Proc. Natl. Acad. Sci.
USA \textbf{95}, 1026 (1998)]. 
\preprint{J. Chem. Phys. 132, 084111 (2010) \hspace{3cm}}
\end{abstract}
\maketitle

\section{Introduction}
The elasticity and configuration of lipid vesicles have attracted
much theoretical attention of physicists
\cite{Lipowsky91,Seifert97,oybook,tuoyjctn2008} since Helfrich
proposed the spontaneous curvature model of lipid bilayers in his
seminal work \cite{helfrich}. The shape equation to describe
equilibrium configurations of lipid vesicles was derived in 1987
\cite{oy87,oy89} based on Helfrich's model. There are two typical
analytical solutions to the shape equations. One is a torus with a
ratio ($\sqrt{2}$) of its two generation radii
\cite{oytorus,Seiferttorus}. Another is a vesicle with biconcave
discoidal shape \cite{Naitooy}. In fact, the latter solution does not correspond to a vesicle free of
external force because of a logarithmic singularity in the solution
\cite{Podgornikpre95,Guvenjpa07,Guvenpre07}.

The opening-up process of lipid vesicles by talin was observed by
Saitoh \emph{et al.} \cite{Hotani}, which pushes us to investigate the
shape equation and boundary conditions of lipid membranes with free
exposed edges. This topic was discussed theoretically and
numerically by several researchers
\cite{Capovilla,Capovilla2,tzcpre,tzcjpa04,yinyjjmb,WangDu06,Umeda05}.
Based on Helfrich's model, the shape equation and boundary
conditions were derived by Capovilla \emph{et al.}
\cite{Capovilla,Capovilla2}, Tu \emph{et al.}
\cite{tzcpre,tzcjpa04}, and Yin \emph{et al.} \cite{yinyjjmb} in
different forms. Wang and Du obtained various shapes of open
membranes through numerical simulations by phase field method
\cite{WangDu06}. Using the area difference elasticity model, Umeda
\emph{et al.} derived the shape equation and boundary conditions
and then compared their numerical results with the experiment
\cite{Umeda05}. They found that the line tension of the free edge of
the open lipid membrane increases with decreasing the concentration
of talin \cite{Umeda05}. The above theoretical and numerical results
can be generalized to investigate adhesions between lipid vesicles
\cite{DesernoCM07,Lv2009}, configurations of lipid vesicles with
different lipid domains
\cite{Lipowskypre96,tzcjpa04,WangDu06,Baumgart05}, and vesicle formation \cite{WangHeJCP09}. However, the above theoretical researches \cite{Capovilla,Capovilla2,tzcpre,tzcjpa04,yinyjjmb} do not contain sufficient discussions on analytical solutions to the shape equation with the
boundary conditions. Additionally, the authors merely compared their numerical results with the experimental ones qualitatively in their numerical work \cite{WangDu06,Umeda05}. It is still lack of the quantitative comparison between the numerical and experimental results. Two natural questions are led to: Can we find analytical solutions? At least, it is instructive to investigate the possibility of finding the analytical solutions. Can we use the numerical results to fit the experimental data quantitatively? We hope we can do that by taking the number of parameters as small as possible.

Generally speaking, the shape equation derived from Helfrich's model
is a fourth-order nonlinear differential equation, while the
boundary conditions include three nonlinear equations describing the
shapes of the free edges of lipid membranes. In principle, one can
obtain the general solution with unknown constants to a linear
differential equation, and then determine the unknown constants by
using the linear boundary conditions. Thus there is no mathematical
difficulty to find the solution satisfying both the linear
differential equation and linear boundary conditions. However, the
problem becomes more complicated if both the differential equation
and boundary conditions are nonlinear. There is no general solution
to a nonlinear differential equation in mathematics. Consequently,
one can only conjecture some special solutions in a few cases. If we
further consider the boundary conditions, only a few ones among the
above known solutions can fit them. Therefore, it is quite helpful
to investigate the feature of the special solutions that can satisfy
both the nonlinear differential equation and the boundary
conditions. Since it is very difficult to obtain solutions to the
shape equation with the boundary conditions, we may first conjecture
a surface satisfying the shape equation, and then find a curve in
the surface satisfying the boundary conditions as an edge of the
surface. However, one might not find any curve satisfying the
boundary conditions for a given surface satisfying the shape
equation. Only some special ones among the surfaces satisfying the
shape equation can admit the boundary conditions. The profound
reason is that the points in the boundary curve should satisfy not
only the boundary conditions, but also the shape equation because
they also locate in the surface. In other words, there exist some
additional constraints between the shape equation and the boundary
conditions. These constraints which have not been touched in
Refs.~\cite{Capovilla,Capovilla2,tzcpre,tzcjpa04,yinyjjmb,WangDu06,Umeda05}
are called the compatibility condition in this paper.

It is not a straightforward task to find the compatibility condition
in general case. The axisymmetric lipid membranes with edges will
give us some clues. The shape equation is reduced to a third-order
differential equation in axisymmetric case
\cite{seifertpra91,jghupre93}. Zheng and Liu proved that it was
integrable \cite{zhengliu93}, and could be further transformed into
a second-order differential equation with an integral constant. In
this paper, we will show that the compatibility condition is that
this integral constant vanishes for axisymmetric membranes. Due to
this compatibility, the shape equation is reduced to a second-order
differential equation while the boundary conditions are reduced to
two equations. The rest of this paper is organized as follows: In
Sec.~\ref{shap-bcs}, we present the general shape equation and
boundary conditions of lipid membranes with free edges. In
Sec.~\ref{sec-const}, we discuss the compatibility between the shape
equation and boundary conditions in axisymmetric case, and then
verify two theorems of non-existence. In Sec.\ref{sec-axisym}, we
find some axisymmetric numerical solutions and compare them with
experimental data quantitatively. A brief summary is given in the
last section.

\section{Shape equation and boundary conditions \label{shap-bcs}}
\begin{figure}[pth!]
\includegraphics[width=7.5cm]{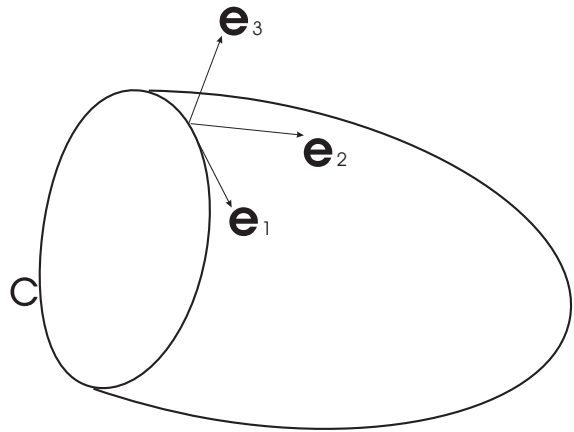}\caption{\label{figframe}
Right-handed frame of an open surface with a boundary curve $C$.}
\end{figure}

A lipid membrane with a free edge is represented as an open surface
with a boundary curve $C$. As shown in Fig.~\ref{figframe}, we can
construct an orthogonal right-handed frame $\{\mathbf{e}_1
,\mathbf{e}_2 ,\mathbf{e}_3\}$ at each point of the surface such
that $\mathbf{e}_3$ is the normal vector of the surface. For each
point in the boundary curve, we take $\mathbf{e}_2$ to be
perpendicular to the tangent direction $\mathbf{e}_1$ of the
boundary curve and point at the side that the surface is located.

The free energy of the membrane may be expressed as
\begin{equation}F=\int [(k_c/2)(2H+c_0)^2 +\bar{k}K] dA + \lambda A +\gamma L, \label{eq-frenergy}\end{equation}
where the first and second terms are the bending energy
\cite{helfrich} and the surface energy of the membrane,
respectively, while the third term is the line energy of the free
exposed edge. $H$ and $K$ are the mean curvature and gaussian
curvature of the surface, respectively. $dA$ is the area element of
the surface. $A$ and $L$ are the total area of the surface and the
total length of the boundary curve, respectively. $k_c$ and
$\bar{k}$ are the bending moduli. $c_0$ is the spontaneous
curvature. $\lambda$ and $\gamma$ are the surface tension and line
tension, respectively.

By calculating the variation of free energy (\ref{eq-frenergy}), we
can obtain \cite{tzcpre}
\begin{equation}(2H+c_{0})(2H^{2}-c_{0}H-2K)-2\tilde\lambda H+\nabla
^{2}(2H) =0, \label{eq-shape}\end{equation} and
\begin{eqnarray}
&&\left. \lbrack (2H+c_{0})+\tilde{k}\kappa_n]\right\vert _{C} =0,\label{bound1} \\
&&\left. \lbrack -2{\partial H}/{\partial\mathbf{e}_2}+\tilde\gamma
\kappa_n+\tilde{k} {d\tau_g}/{ds}]\right\vert _{C} =0,\label{bound2}\\
&&\left. \lbrack (1/{2})(2H+c_{0})^{2}+\tilde{k}K+\tilde\lambda
+\tilde\gamma \kappa_{g}]\right\vert _{C}=0,\label{bound3}
\end{eqnarray}
where $\tilde{\lambda}\equiv\lambda/k_c$,
$\tilde{k}\equiv\bar{k}/k_c$, $\tilde{\gamma}\equiv\gamma/k_c$ are the reduced surface tension, reduced bending modulus, and reduced line tension, respectively. $\kappa_n$, $\kappa_g$, $\tau_g$, and $ds$ are the normal
curvature, geodesic curvature, geodesic torsion, and arc length
element of the boundary curve, respectively. Equation
(\ref{eq-shape}) determines the equilibrium shape of the membrane,
thus we call it shape equation. For a given surface satisfying the
shape equation, Eqs.~(\ref{bound1})-(\ref{bound3}) determine the
shape of the boundary curve and its position in the surface, thus we
call them boundary conditions. Equation~(\ref{eq-shape}) expresses
the normal force balance of the membrane. Equation (\ref{bound1}) is
the moment balance equation around $\mathbf{e}_1$ at each point in
curve $C$. Equations~(\ref{bound2}) and (\ref{bound3}) are the force
balance equations along $\mathbf{e}_3$ and $\mathbf{e}_2$ at each
point in curve $C$, respectively. Thus, in general, the above four
equations are independent of each other.

\section{compatibility between the shape equation and boundary conditions\label{sec-const}}
We have mentioned that only some special ones among the surfaces
satisfying the shape equation (\ref{eq-shape}) can admit the
boundary conditions (\ref{bound1})-(\ref{bound3}). What is the
common feature of these special surfaces? we will find this feature
for axisymmetric surfaces.

\begin{figure}[pth!]
\includegraphics[width=7.5cm]{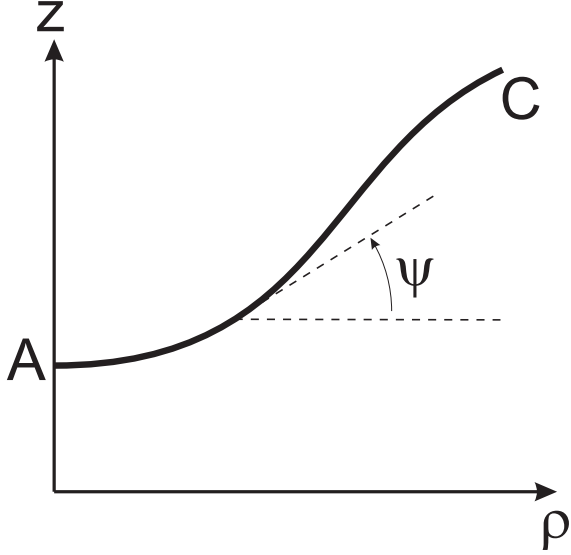}\caption{\label{figoutline}
Outline of an open surface. Each open surface can be generated by a
planar curve AC rotating around z axis. $\psi$ is the angle between
the tangent line and the horizontal plane.}
\end{figure}

When a planar curve AC shown in Fig.\ref{figoutline} revolves around
$z$ axis, an axisymmetric surface is generated. Let $\psi$ represent
the angle between the tangent line and the horizontal plane. Each
point in the surface can be expressed as vector form
$\mathbf{r}=\{\rho\cos \phi,\rho\sin \phi,z(\rho)\}$ where $\rho$
and $\phi$ are radius and azimuth angle that the point corresponds
to. Introduce a notation $\sigma$ such that $\sigma =1$ if
$\mathbf{e}_1$ is parallel to $\partial\mathbf{r} /\partial \phi$,
and $\sigma =-1$ if $\mathbf{e}_1$ is antiparallel to
$\partial\mathbf{r} /\partial \phi$ in the boundary curve generated
by point C. The above equations (\ref{eq-shape})-(\ref{bound3}) are
transformed into
\begin{eqnarray}
(h-c_{0})\left(\frac{h^{2}}{2}+\frac{c_{0}h}{2}-2K\right)-\tilde{\lambda}
h+\frac{\cos \psi }{\rho}(\rho\cos \psi h')'=0,\label{sequilib}
\\
\left[h-c_{0}+\tilde{k}{\sin \psi }/{\rho}\right]_C=0,\label{sbound1}\\
\left[-\sigma\cos \psi h'+\tilde{\gamma}{\sin \psi
}/{\rho}\right]_C=0,\label{sbound2}\\
\left[\frac{\tilde{k}^2}{2}\left(\frac{\sin \psi
}{\rho}\right)^2+\tilde{k}K+\tilde{\lambda}-\sigma\tilde{\gamma}
\frac{\cos \psi }{\rho}\right]_C=0,\label{sbound3}
\end{eqnarray}
where $h\equiv {\sin \psi }/{\rho}+(\sin\psi)'$ and $K\equiv{\sin \psi
}(\sin\psi)'/{\rho}$. The `prime' represents the derivative with
respect to $\rho$.

The shape equation (\ref{sequilib}) is a third-order differential
equation. Following Zheng and Liu's work \cite{zhengliu93}, we can
transform it into a second order differential equation
\begin{eqnarray}\cos\psi h'
&+&(h-c_{0}) \sin\psi\psi^{\prime}
  -\tilde{\lambda} \tan\psi\nonumber\\&+&\frac{\eta_{0}}{\rho\cos\psi}-\frac{\tan\psi}%
{2}(h-c_{0})^{2} =0\label{firstintg}\end{eqnarray} with an integral
constant $\eta_{0}$ (so called the first integral). The
configuration of an axisymmetric open lipid membrane should satisfy
the shape equation (\ref{sequilib}) or (\ref{firstintg}) and
boundary conditions (\ref{sbound1})-(\ref{sbound3}). In particular,
the points in the boundary curve should satisfy not only the
boundary conditions, but also the shape equation (\ref{firstintg})
because they also locate in the surface. That is,
Eqs.~(\ref{sbound1})-(\ref{sbound3}) and (\ref{firstintg}) should be
compatible with each other in the edge. Substituting
Eqs.~(\ref{sbound1})-(\ref{sbound3}) into (\ref{firstintg}), we
derive the compatibility condition to be
\begin{equation}\eta_{0}=0.\label{const-condit}\end{equation}

No we will discuss two examples and verify two theorems of
non-existence by considering the above compatibility condition.

\begin{figure}[pth!]
\includegraphics[width=7.5cm]{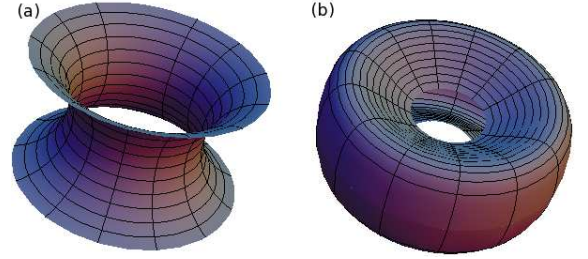}\caption{(Color online)\label{figtorus}
Two non-existent axisymmetric open membranes: (a) A part of a torus; (b) A part of biconcave discodal surface.}
\end{figure}

First, let us consider a part of a torus shown in Fig.~\ref{figtorus}a generated by an arc expressed by
$\sin\psi=\alpha\rho+\beta$ with two non-vanishing constants
$\alpha$ and $\beta$. Substituting it into the shape equation
(\ref{firstintg}), we obtain $c_0 =0$, $\beta=\sqrt{2}$,
$\tilde{\lambda}=0$, and $\eta_0 = - \alpha$. That is, the torus can
be a solution to the shape equation. However, $\eta_0 = - \alpha
\neq 0$ contradicts to the compatibility condition
(\ref{const-condit}). Thus we arrive at:

\emph{Theorem 1}. There is no axisymmetric open membrane being a
part of torus generated by a circle expressed by
$\sin\psi=\alpha\rho+\sqrt{2}$.

Secondly, we consider a biconcave discodal surface \cite{Naitooy}
generated by a planar curve expressed by $\sin\psi=\alpha \rho
\ln(\rho/\beta)$ with two non-vanishing constants $\alpha$ and
$\beta$. To avoid the logarithmic singularity at two poles,
we may dig two holes around the poles in the surface as shown in Fig.~\ref{figtorus}b.
Substituting $\sin\psi=\alpha \rho
\ln(\rho/\beta)$ into the shape equation (\ref{firstintg}),
we obtain $\tilde{\lambda}=0$, $\alpha=c_0$, and $\eta_0 = -2 c_0$.
That is, the biconcave discodal surface can be a solution to the
shape equation. However, $\eta_0 = - 2c_0 \neq 0$ contradicts to the compatibility
condition (\ref{const-condit}). Thus we arrive at:

\emph{Theorem 2}. There is no axisymmetric open membrane being a
part of a biconcave discodal surface generated by a planar curve
expressed by $\sin\psi=c_0 \rho \ln(\rho/\beta)$.

In the above discussion, the theorems of non-existence are deduced as natural corollaries of the compatibility condition. It does not mean that the proofs are unique. The other proofs are presented in Appendix~\ref{app-proof}. In Ref.~\cite{tzc09rome}, the present author has proved that there is no open lipid membrane being a part of a constant mean curvature surface. These theorems reveal that it is almost hopeless to find analytical solutions to the shape equations with the boundary conditions. Thus we need to seek for numerical solutions.

\section{Axisymmetric numerical solutions\label{sec-axisym}}
The compatibility condition leads to a more important result that
the shape equation can be simplified as
\begin{eqnarray}\cos\psi h'
+(h-c_{0}) \sin\psi\psi^{\prime}
  -\tilde{\lambda} \tan\psi-\frac{\tan\psi}%
{2}(h-c_{0})^{2} =0,\label{newshapeq}\end{eqnarray} while the
boundary conditions can be taken only two equations (\ref{sbound1})
and (\ref{sbound3}) because Eq.~(\ref{sbound2}) is not independent
of Eqs.~(\ref{sbound1}), (\ref{sbound3}), and (\ref{newshapeq}).
However, it is still very difficult to obtain analytical solutions to
Eq.(\ref{newshapeq}) with boundary conditions (\ref{sbound1}) and
(\ref{sbound3}). We will find axisymmetric numerical solutions and
compare them with experimental data \cite{Hotani} in this section.

Because $\psi$ might be a multi-valued function of the independent
variable $\rho$, the above equations are unsuitable for numerical
solutions. Here we take the arc-length of curve AC in
Fig.~\ref{figoutline} as an independent variable. Then we have
$\dot{\rho}=\cos\psi$ and $\dot{z}=\sin\psi$, where the `dot'
represents the derivative with respect to the arc-length. The shape
equation can be transformed into
\begin{equation}\ddot{\psi}=-\frac{\tan\psi}{2}\dot{\psi}^{2}-\frac{\cos\psi\dot{\psi}}%
{\rho}+\frac{\sin2\psi}{2\rho^{2}}+\tilde{\lambda}\tan\psi+\frac{\tan\psi}{2}\left(
\frac{\sin\psi}{\rho}-c_{0}\right) ^{2},\label{neweq2}\end{equation}
while the boundary conditions become
\begin{equation}\left[\dot{\psi}-c_{0}+\left(  1+\tilde{k}\right)  \frac{\sin\psi}{\rho}\right]_C=0,\label{newbcs1}\end{equation}
and \begin{equation}\left[\tilde{k}c_{0}\frac{\sin\psi}{\rho}-\left(
1+\frac{\tilde{k}}{2}\right) \tilde{k}\left(
\frac{\sin\psi}{\rho}\right)  ^{2}+\tilde{\lambda}+\tilde{\gamma
}\frac{\cos\psi}{\rho}\right]_C  =0.\label{newbcs2}\end{equation} In
fact, these equations can be also derived from the Lagrangian method
as shown in Appendix~\ref{app-deriv}. In addition, we impose the
initial conditions $z(0)=\rho(0)=0~\mu$m, and $\psi(0)=0$. We can
use the shooting method to find numerical solutions to
Eq.~(\ref{neweq2}) with boundary conditions [Eqs.~(\ref{newbcs1})
and (\ref{newbcs2})] and these initial conditions, and then fit the
parameters ($\tilde{k}$, $c_0$, $\tilde\lambda$, $\tilde\gamma$)
with experimental data. The basic idea is as below. For the given
values of $\tilde{k}$, $c_0$, $\tilde\lambda$, $\tilde\gamma$ and
$\dot{\psi}(0)$, we can solve Eq.~(\ref{neweq2}) with boundary
conditions (\ref{newbcs1}) and (\ref{newbcs2}). Then we compare the
graph of the solution to the outline of the open membrane in the
experiment. Tune the values of the parameters until the graph of the
solution and the outline of the membrane almost superpose each
other. Thus we obtain a group of proper values of the parameters
($\tilde{k}$, $c_0$, $\tilde\lambda$, $\tilde\gamma$).

\begin{figure}[pth!]
\includegraphics[width=7.7cm]{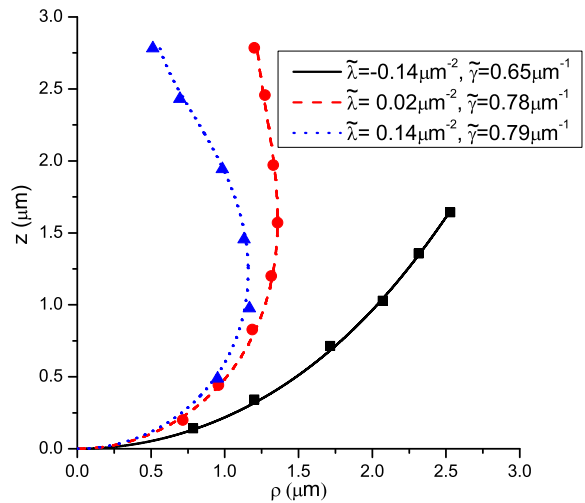}\caption{\label{fig3i-k}(Color online) Numerical results (solid, dash, and
dot lines) and experimental data (squares, circles, and triangles
extracted from Fig.~3.I to K in Ref.~\cite{Hotani}) of the outlines
of an axisymmetric lipid membrane at different concentration of
talin.}
\end{figure}

In the experiment \cite{Hotani}, the hole of the lipid membrane is
enlarged with increasing the concentration of talin, and vice versa.
Talin molecules adhere to the edge of the membrane. Thus it is
reasonable to assume that the line tension of the edge depends on
the concentration of talin, while the bending moduli and spontaneous
curvature of the membrane do not. That is, we should have the common
values of $\tilde{k}$ and $c_0$ for a membrane at different
concentrations of talin. This gives a constraint in our fitting. As
shown in Fig.~\ref{fig3i-k}, our numerical results (solid, dash, and
dot lines) obtained from Eqs.~(\ref{neweq2})-(\ref{newbcs2}) agree
well with the experimental data (squares, circles, and triangles
extracted from the outlines of the membrane with decreasing the
concentration of talin). The common parameters are fitted as
$\tilde{k}= -0.122$ and $c_0 = 0.4~\mu$m$^{-1}$. The negative
$\tilde{k}$ reveals that the surface with positive Gaussian
curvature is more energetically favorable than that with negative
Gaussian curvature. The positive $c_0$ reflects the asymmetry of the
bilayer composed in the experiment \cite{Hotani}, which makes the
membrane bend like a standing upright cup. The other parameters are
shown in the figure. The reduced line tension $\tilde\gamma$
increases from 0.66~$\mu$m$^{-1}$ to 0.79~$\mu$m$^{-1}$ with
decreasing the concentration of talin, which is the same as the
conclusion of Ref.~\cite{Umeda05}. The surface tension depends on
the shapes and line tension. Intuitively, the line tension of the
edge induces compression stress in the membranes with the similar
shapes generated by solid line in Fig.~\ref{fig3i-k}, thus the
surface tension is negative. On the contrary, the tension of the
edge induces stretching stress in the membranes with the similar
shapes generated by dash or dot lines in Fig.~\ref{fig3i-k}, thus
the surface tension is positive. The variation of surface tension
can be comprehensively understood from Eq.~(\ref{constraitg}). For
the membrane in Fig.~\ref{fig3i-k},
$2H=-({\sin\psi}/{\rho}+\dot{\psi})<0$ and decreases according to
the sequence of the solid, dash and dot lines. Therefore, the
surface tension of the membrane increases according to the same
sequence with considering Eq.~(\ref{constraitg}). Furthermore, we
also examine that our numerical results indeed satisfy the
constraint (\ref{constraitg}).

\section{Conclusion}
In the above discussion, we investigate the compatibility between
shape equation and boundary conditions of lipid membranes with free
edges. The main results obtained in this paper are as follows.

(i) The compatibility condition for axisymmetric lipid membranes with free
edges is that the first integral of the shape equation
(\ref{sequilib}) should be vanishing, i.e.,
Eq.~(\ref{const-condit}).

(ii) Two theorems (\emph{Theorem 1} and \emph{Theorem 2} in
Sec.~\ref{sec-const}) of non-existence are verified as natural corollaries of the compatibility condition, which give two
examples to reveal that one indeed might not find any curve satisfying
the boundary conditions in a given surface satisfying the shape
equation. These theorems also correct two flaws on analytical solutions in Ref.~\cite{tzcpre}.

(iii) The shape equation of axisymmetric lipid membranes is reduced
to Eq.(\ref{newshapeq}). Then only two equations in boundary
conditions are independent. This conclusion is the same as the case in
Ref.~\cite{Capovilla} with vanishing $\bar{k}$ and $c_0$.

(iv) As shown in Fig.~\ref{fig3i-k}, the numerical solutions to the
reduced shape equation (\ref{newshapeq}) with boundary conditions
(\ref{sbound1}) and (\ref{sbound3}) agree well with the experimental
data \cite{Hotani}.

Finally, we would like to point out two difficulties that we have
not fully overcome yet: (i) The compatibility condition between
shape equation and boundary conditions for asymmetric (not
axisymmetric) lipid membranes with edges is unclear. We do not even
know whether it exists, much less what it is. (ii) We use the
shooting method to find numerical solutions. But this method is not
so efficient to the numerical solutions due to the complicated
boundary conditions. A much more efficient method is expected. The
above challenges should be addressed in the future work.

\section*{Acknowledgement}
The author is grateful to X. H. Zhou and M. Li for their instructive
discussions, and to Nature Science Foundation of China (grant no.
10704009) and the Foundation of National Excellent Doctoral
Dissertation of China (grant no. 2007B17) for their financial supports.

\appendix
\section{Other proofs to theorems non-existence\label{app-proof}}
The proofs can be divided into two classes in terms of different starting points. One is based on the stress analysis, another is based on the scaling argument.
\subsection{Stress analysis}
Capovilla \emph{et al.} proposed the stress tensor in a lipid membrane and then derived the shape equation and boundary conditions from the stress tensor \cite{Capovilla,Capovilla2}. Recently, they found that \cite{Guvenjpa07,Guvenpre07} the line integral
\begin{equation}\oint_\Gamma ds l^a \mathbf{f}_a \cdot \hat{z}=c,\end{equation}
where $\Gamma$ is any circle perpendicular to the symmetric axis in an axisymmetric membrane. $l^a$ and $\mathbf{f}_a$ represent the normal of $\Gamma$ tangent to the membrane surface and the stress in the membrane, respectively. $\hat{z}$ is the unit vector along the symmetric axis. $c$ is a constant dependent on the topology and the curvature singularity of the membrane.

First, the constant $c$ is non-vanishing for an axisymmetric torus free of curvature singularity, which implies that the stress in each circle $\Gamma$ perpendicular to the symmetric axis in the torus surface cannot be zero. However, the stress in the free edges should be vanishing. Thus, we cannot find any $\Gamma$ as a free edge of an axisymmetric open membrane being a
part of the torus, i.e., theorem 1 is arrived at.

Secondly, there exist singularity points at two poles of the biconcave discodal surface generated by a planar curve expressed by $\sin\psi=\alpha \rho
\ln(\rho/\beta)$. The singularity results in a non-vanishing $c$ \cite{Guvenjpa07,Guvenpre07}, and then non-zero stress in each circle $\Gamma$ in the biconcave discodal surface. Thus, we cannot find any $\Gamma$ as a free edge of an axisymmetric open membrane being a
part of the biconcave discodal surface, i.e., theorem 2 is arrived at.

\subsection{Scaling argument}
The free energy (\ref{eq-frenergy}) can be written in another form
\begin{eqnarray}F&=&\int [(k_c/2)(2H)^2 +\bar{k}K] dA\nonumber\\ &+& 2k_c c_0\int H dA+(\lambda+k_c c_0^2 /2) A +\gamma L. \label{eq-frenergyn2}\end{eqnarray}
Let us consider the scaling transformation $\mathbf{r}\rightarrow \Lambda\mathbf{r}$, where the vector $\mathbf{r}$ represents the position of each point in the membrane and $\Lambda$ is a scaling parameter \cite{Capovilla2}. Under this transformation, we have $A\rightarrow \Lambda^2 A$, $L\rightarrow \Lambda L$, $H\rightarrow \Lambda^{-1} H$, and $K\rightarrow \Lambda^{-2} K$. Thus, Eq.~(\ref{eq-frenergyn2}) is transformed into
\begin{eqnarray}F(\Lambda)&=&\int [(k_c/2)(2H)^2 +\bar{k}K] dA\nonumber\\ &+& 2k_c c_0 \Lambda \int H dA+(\lambda+k_c c_0^2 /2)\Lambda^2 A +\gamma \Lambda L. \label{eq-frenergyn3}\end{eqnarray}

The equilibrium configuration should satisfy $\partial F/\partial\Lambda =0$ when $\Lambda=1$ \cite{Capovilla2}. Thus we obtain
\begin{equation}2 c_0 \int H dA+(2\tilde\lambda+c_0^2) A +\tilde\gamma L=0.\label{constraitg}\end{equation}
This equation is an additional constraint for open membranes.

As shown in Sec.~\ref{sec-const}, if there exists an open membrane being a part of torus, then the shape equation (\ref{eq-shape}) requires $\tilde\lambda =0$ and $c_0 =0$, which contradicts the constraint (\ref{constraitg}) because $\tilde\gamma L>0$. Thus we arrive at theorem 1.

Because Willmore surfaces satisfy the special form of Eq.~(\ref{eq-shape}) with vanishing $\tilde\lambda$ and $c_0$ \cite{Willmore82}, as a byproduct of the constraint (\ref{constraitg}), we obtain a much stronger theorem of non-existence: There is no open membrane being a
part of a Willmore surface.

Next, we turn to the biconcave discodal surface. If there exists an open membrane being a
part of a biconcave discodal surface generated by a planar curve
expressed by $\sin\psi=c_0 \rho \ln(\rho/\beta)$, the shape equation (\ref{eq-shape}) requires $\tilde\lambda =0$. Substituting $2H=-c_0[1+2 \ln(\rho/\beta)]$ into Eq.~(\ref{constraitg}), we will not obtain a contradiction. Thus theorem 2 cannot be deduced from the scaling argument.

\section{Derivation of the reduced shape equation and boundary conditions by using the Lagrange method \label{app-deriv}}
For the revolving surface generated by the planar curve shown in
Fig.~\ref{figoutline}, Eq.~(\ref{eq-frenergy}) can be transformed
into
\begin{equation}{F}/{2\pi
k_c}=\int_0^{s_2}[\rho
f^{2}/2+\tilde{k}\sin\psi\dot{\psi}+\tilde{\lambda} \rho
+\tilde{\gamma} \dot{\rho}]ds,\end{equation} with
$f={\sin\psi}/{\rho} +\dot{\psi}-c_{0}$. We should minimize
${F}/{2\pi k_c}$ with the constraints $\dot{\rho}=\cos\psi$ and
$\dot{z}=\sin\psi$, thus we construct an action $S=\int_0^{s_2}
\mathcal{L} ds$ with a Lagrangian \cite{oybook}
\begin{equation}\mathcal{L}=\rho f^{2}/2+\tilde{k}\sin\psi\dot{\psi}+\tilde{\lambda}
\rho +\tilde{\gamma}
\dot{\rho}+\zeta(\dot{\rho}-\cos\psi)+\eta(\dot{z}-\sin\psi),\label{lagrangL}\end{equation}
where $\zeta$ and $\eta$ are two Lagrange multipliers. In terms of
the variational theory, we can derive
\begin{eqnarray}
\delta S  &  =&\int_{0}^{s_{2}}\delta \mathcal{L}ds-\mathcal{H}\delta s_{2}\nonumber\\
&=&\int_{0}^{s_{2}}\left(  \frac{\partial \mathcal{L}}{\partial\psi}-\frac{d}%
{ds}\frac{\partial \mathcal{L}}{\partial\dot{\psi}}\right)
\delta\psi ds+\left.
\frac{\partial \mathcal{L}}{\partial\dot{\psi}}\delta\psi\right\vert _{0}^{s_2}\nonumber\\
&+&\int_{0}^{s_{2}}\left(  \frac{\partial \mathcal{L}}{\partial \rho}-\frac{d}%
{ds}\frac{\partial \mathcal{L}}{\partial\dot{\rho}}\right)  \delta
\rho ds+\left.
\frac{\partial \mathcal{L}}{\partial\dot{\rho}}\delta \rho\right\vert _{0}^{s_2}\nonumber\\
&+&\int_{0}^{s_{2}}\left(  0-\frac{d}{ds}\frac{\partial
\mathcal{L}}{\partial \dot{z}}\right)  \delta zds+\left.
\frac{\partial \mathcal{L}}{\partial\dot{z}}\delta
z\right\vert _{0}^{s_2}\nonumber\\
&+&\int_{0}^{s_{2}}\left(  \dot{\rho}-\cos\psi\right)  \delta\zeta
ds+\int_{0}^{s_{2}}\left(  \dot{z}-\sin\psi\right)  \delta\eta ds\nonumber\\
&-&\mathcal{H}\vert_C\delta s_{2} =0,\label{deltaS}
\end{eqnarray}
where the Hamiltonian $\mathcal{H}=\dot{\psi}\frac{\partial
\mathcal{L}}{\partial\dot{\psi}}+\dot{\rho}\frac{\partial
\mathcal{L}}{\partial\dot{\rho}}+\dot{z}\frac{\partial
\mathcal{L}}{\partial\dot{z}}-\mathcal{L}$. Imposing $\psi(0)=0$,
$\rho(0)=z(0)=0~\mu$m, and substituting Eq.~(\ref{lagrangL}) into
Eq.~(\ref{deltaS}), we can obtain
\begin{eqnarray}&&\zeta\sin\psi-\rho \dot{f} =0,\label{tmpeq1}\\
&&\eta=\mathrm{constant},\label{tmpeq2}\\
&&\dot{\rho}=\cos\psi,\\
&&\dot{z}=\sin\psi,
\end{eqnarray}
with boundary conditions
\begin{eqnarray}&&\left[f  +\tilde{k}\frac{\sin\psi}{\rho}\right]_C=0,\label{tmpbc1}\\
&&\zeta\vert _{C}+\tilde{\gamma}=0,\label{tmpbc2}\\
&&\eta\vert_{C}=0,\label{tmpbc3}\\
&&\mathcal{H}\vert_C=0.\label{tmpbc4}
\end{eqnarray}
Eq.~(\ref{tmpbc1}) is equivalent to boundary condition
(\ref{newbcs1}).

Because $\mathcal{L}$ does not explicitly contain $s$, $\mathcal{H}$
is a constant. Combining Eqs.~(\ref{tmpeq2}), (\ref{tmpbc3}),
(\ref{tmpbc4}) and the definition of $\mathcal{H}$, we derive
\begin{equation}\zeta\cos\psi=\rho [f  (f-2\dot{\psi})/2  +\tilde{\lambda} ].\label{tmpeq3}\end{equation}
From Eqs.~(\ref{tmpeq1}) and (\ref{tmpeq3}) we can obtain the shape
equation (\ref{neweq2}). Equation (\ref{tmpbc4}) can be transformed
into the boundary condition (\ref{newbcs2}) with Eq.~(\ref{tmpbc1}).
From Eqs.~(\ref{tmpeq1}) and (\ref{tmpbc2}) we can also obtain the
other boundary condition, which is not independent of the shape
equation (\ref{neweq2}) and boundary conditions (\ref{newbcs1}) and
(\ref{newbcs2}).

\end{document}